\documentclass[a4paper]{article}

\usepackage{INTERSPEECH2021}

\usepackage{url}
\usepackage{hyperref}
\usepackage{graphicx}
\usepackage{amssymb,amsmath,bm}
\usepackage{algorithm2e}
\usepackage{textcomp}
\usepackage{comment}
\usepackage{multirow}
\usepackage{stfloats}
\usepackage{float}

\def\vec#1{\ensuremath{\bm{{#1}}}}

\title{Auto-KWS 2021 Challenge: Task, Datasets, and Baselines}

\makeatletter
\def\name#1{\gdef\@name{#1\\}}
\makeatother

\name{{\em Jingsong Wang$^1$, Yuxuan He$^1$, Chunyu Zhao$^1$, Qijie Shao$^2$, Wei-Wei Tu$^{1,3}$, Tom Ko\thanks{*corresponding author}$^4*$, Hung-yi Lee$^5$, Lei Xie$^2$}} 

\address{$^{1}$4Paradigm Inc.  \\
  $^2$ASLP@NPU, Northwestern Polytechnical University \\ 
  $^3$ChaLearn \\
  $^4$Department of Computer Science and Engineering\\ Southern University of Science and Technology \\
  $^5$College of Electrical Engineering and Computer Science, National Taiwan University
}
\email{wangjingsong@4paradigm.com}

\begin{document}

\maketitle
\begin{abstract}

Auto-KWS 2021 challenge calls for automated machine learning (AutoML) solutions to automate the process of applying machine learning to a customized keyword spotting task. 
Compared with other keyword spotting tasks, Auto-KWS challenge has the following three characteristics: 
1) The challenge focuses on the problem of customized keyword spotting, where the target device can only be awakened by an enrolled speaker with his specified keyword. The speaker can use any language and accent to define his keyword.
2) All dataset of the challenge is recorded in realistic environment. It is to simulate different user scenarios.
3) Auto-KWS is a ``code competition", where participants need to submit AutoML solutions, then the platform automatically runs the enrollment and prediction steps with the submitted code.
This challenge aims at promoting the development of a more personalized and flexible keyword spotting system.
Two baseline systems are provided to all participants as references.
\end{abstract}

\section{Introduction}
\label{sect:intro}
\vspace{2mm}
Recently, the keyword spotting (KWS) technology\cite{chen2014small,zhang2017hello,tang2018deep,shan2018attention} has started emerging in people's daily life through smart speakers and in-vehicle devices.
Well-known examples of KWS applications include ``OK Google" in Google Home and ``Hey Siri" in Apple’s Siri. People can wake up the device by speaking the predefined keyword.
Meanwhile, the KWS task is extended to consider the security. 
Personalized wake-up mode, including customized wake-up word detection and specific voiceprint verification, has created more application scenarios.
The Auto-KWS challenge, which is an automated machine learning challenge for customized keyword spotting, is designed to address this difficult problem. 
In this challenge, participants need to design a computer program to solve the customized keyword spotting problem autonomously (without any human intervention).

The Auto-KWS challenge simulates the scenario where a device can be customized for its wake-up condition.
For a new enrollment, the user selects a keyword and reads it repeatedly. The wake-up device records the content and voiceprint of these sound clips. In the later stage, the device determines whether it is wakened up by comparing the input voice with the enrollment information. Some conventional wake-up devices also support multiple keywords or voiceprint authentication. However, wake-up word detection and speaker verification (SV) are carried out separately in the pipeline, where a wake-up word detection system is used to generate a successful trigger, followed by a speaker verification system used to perform identity authentication\cite{jia20212020}.
Besides that, both the keyword spotting and voiceprint verification are text-dependent, limiting the flexibility to customise wake-up words.

Machine learning, especially deep learning, has brought great improvements to the field of speech technology related to this challenge.
Deep neural networks(DNN) based KWS have gradually replaced conventional approaches, and more complex network structures have been adopted to promote the performance of KWS systems\cite{sainath2015convolutional,tang2018deep,wang2020wake}.
Moreover, query by example (QbE) methods, which compare the test audio segment against the templates to make detection decisions, can improve the KWS performance\cite{chen2015query,yuan2018learning,yuan2019verifying}.
On the other hand, DNNs have also shown great performance on speaker verification tasks \cite{rahman2018attention,ko2020prototypical}.

There are some prior works on combining keyword detection and speaker verification\cite{jia20212020,hou2021npu}, for example, a text-dependent speaker verification task with fixed keywords.
Their approaches combine two independently trained sub-systems and present a joint end-to-end neural network system.
Meanwhile, the series of pretrained models offer vector representations, including pretraining with generative loss~\cite{audio-word2vec, apc1, mockingjay} or discriminative loss~\cite{cpc, wav2vec}. 
Audio Word2vec\cite{audio-word2vec} extracted fixed-dimension segmental representation for query-by-example; \cite{apc1, mockingjay, pase} demonstrated a sequence of extracted representations for entire utterances can capture phonetic, speaker, or emotion characteristics; \cite{wav2vec, vq_wav2vec, wav2vec2, decoar2} showed that SSL can establish new principles to solve ASR problem.

The solution to this challenge may contain a complex pipeline. Automatic machine learning (AutoML) related technologies or methods can help build a solution system which has good performance and high computational efficiency \cite{yao2018taking}, such as proposing more effective representation for enrollment utterances of each speaker, or a more robust and effective feature-based method than DTW method \cite{giorgino2009computing}. In low resource situation, meta-learning or few shot learning have the ability to fully utilize limited data and improve the performance of the KWS or SV model under difficult conditions \cite{liu2020metadata,parnami2020few,chen2018investigation,ko2020prototypical}.

The Auto-KWS Challenge is the third in a series of automated speech challenges\footnote{https://www.4paradigm.com/competition/autospeech2020}\footnote{https://www.4paradigm.com/competition/autospeech2021}, which applies AutoML to the tasks in speech processing\cite{wang2020autospeech}.
Unlike many other challenges \cite{jia20212020,fu2020ieee}, we require code submission instead of prediction submission.  
Participants’ code will be automatically run on multiple datasets on competition servers with the same hardwares (CPU, GPU, RAM, Etc.) in order to have fair comparisons. 
As the test datasets on the platform are unseen by the participants, we provide the practice data, in order to facilitate the participants' offline debugging. 
The submitted code should complete the enrollment, prediction and other processes within a specified time budget. The platform will calculate a score according to the predictions on the test datasets.

The paper is organized as follows: Section 2 describes the design of the competition, including competition protocol, metrics, datasets and starting kit. Section 3 describes two baselines we use and the results of the experiments. Section 4 presents the conclusions.

\section{Competition Design}
\label{sec:competition-design}
\vspace{4mm}

\subsection{Competition protocol}
\label{subsec:protocol}
The Auto-KWS 2021 competition has 3 phases: feedback phase, check phase, and final phase. Before the feedback phase, participants are provided with the training and practice datasets to develop their solutions offline. During the feedback phase, participants can upload their solutions to the platform to receive immediate performance feedback for validation. Then in the check phase, the participants can submit their code only once to make sure their code works properly on the platform. The platform will indicate success or failure to the participants, but detailed logs will be hidden. Lastly, in the final phase, participants' solutions will be evaluated on the private dataset. Once the participants submit their code, the platform will run their algorithms automatically to test on the private data with time budget. The final ranking will be generated based on the scores calculated in this phase.

The platform exploits the same evaluation logic in all phases, shown in Figure 1. The task is defined by:
$$\mathcal{T} = (D_{e}, D^{\emptyset}_{te}, L, B_T, B_S)$$ 
where $D_{{e}}$ and $D^{\emptyset}_{te}$ represent examples in the enrollment data and test data without labels respectively, $L: {Y}\times {Y}\to \mathbb{R}$ is a loss function measuring the losses of predictions $y'$ with respect to the true labels $y$. 
$B_S$ is the space budget and $B_T$  contains the time budget for initialization, enrollment and prediction programs. 

The initialization time budget is 30 minutes, the enrollment time budget is 5 minutes for each speaker, and the test time budget is Real Time Factor (RTF) $F_r$ times the total duration of the test audio. $F_r$ is calculated as, 
$$ F_r = T_{\text{process}} / T_{\text{data}}$$
where $T_{\text{process}}$ is the total processing time of all test data, and $T_{\text{data}}$ is the total duration of the test audio. Notice that $T_{\text{process}}$ only includes the inference time for each test audio since the enrollment and initialization process has already been completed. In each task, after initialization, the platform will call the ``enrollment.sh" script, which runs for 5 minutes and then call the ``predict.sh" script to predict the label for each test audio. When $B_T$ or $B_S$ runs up, the platform will automatically terminate the processes.

For each speaker's dataset, we calculate the wake-up score $\text{S}_i$ from the Miss Rate (MR) and the False Alarm Rate (FAR) in the following form:\\
$$ \text{S}_i=MR + \alpha \times \text{ FAR} $$
where $\alpha$ is a variable representing the penalty coefficient and set to 9 in this challenge.
The final score is an average of all the $S_i$.

\begin{figure}
\centering
\includegraphics[scale=.31]{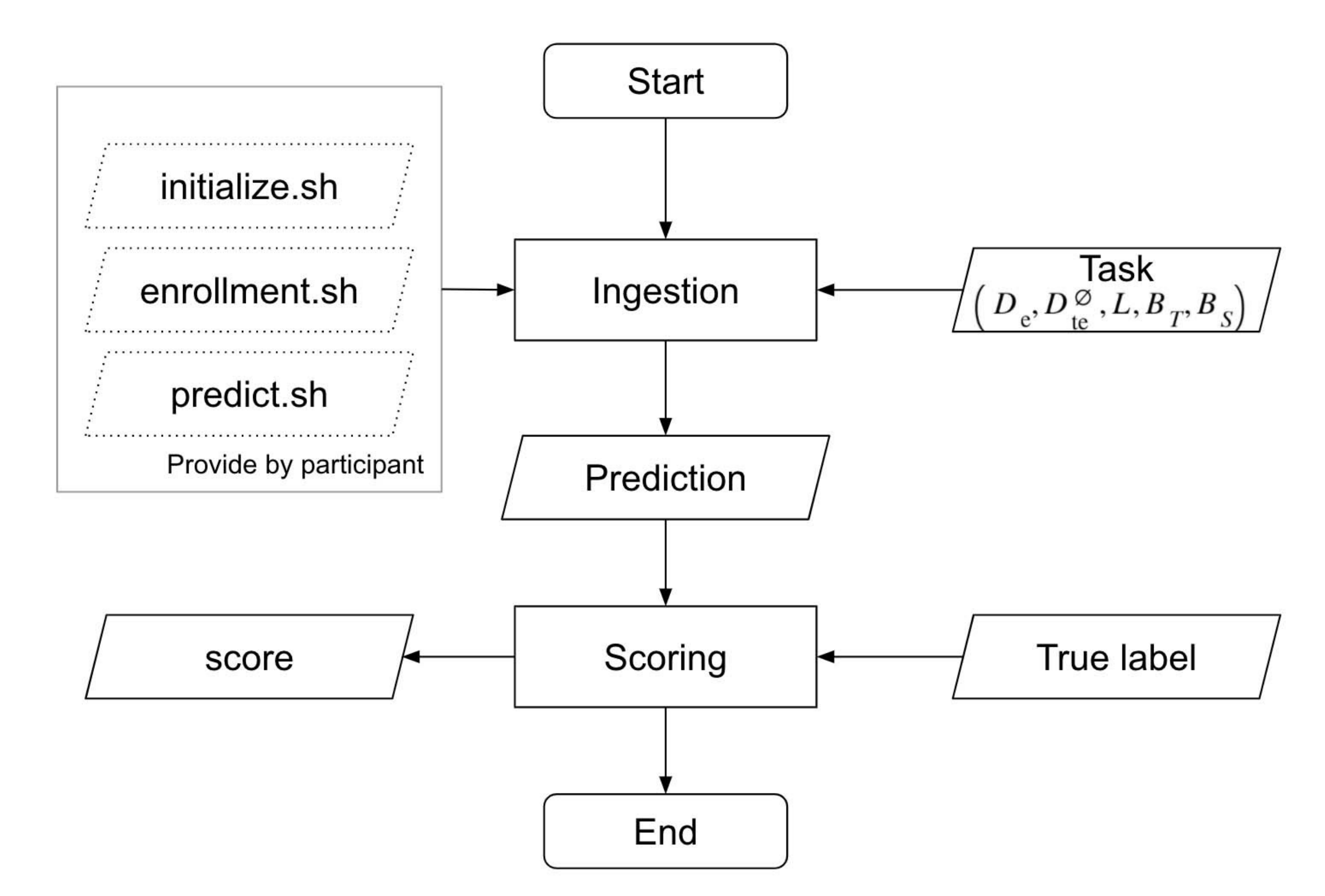}
\caption{Auto-KWS evaluation process for one task. The ingestion program will look for ``initialize.sh", ``enrollment.sh" and ``predict.sh" from participant's submitted code folder and generate prediction results against the hidden task data. The prediction output will be passed into the scoring program to compare against the true label. The ingestion program contains the time and space budget $B_T$, $B_S$ for initialization, enrollment and prediction processes respectively}
\end{figure}

\subsection{Datasets}
\label{subsec:datasets}
An online recording website collected the data used in the Auto-KWS 2021 Challenge. When speakers entered the website, the first thing was to select a language, i.e., Chinese or English. If Chinese was selected, they needed to fill in the dialect they would use. After that, the website would display the corresponding text we provided. Speakers needed to read out the text and uploaded audios recorded by the website to the server. After receiving the audios, we manually screened the data to remove unusable clips. In the end, we got 206 audios read by 165 speakers.

\begin{table}[H]
    \caption{\textit{{\textbf{Language/dialect summary.} The contributors to voices in the dataset are randomly selected. They come from many different regions of China. Therefore, our dataset's accents are diverse, and one quarter of them are dialects, including Cantonese, Szechwanese and many others.}}}
    \centering
    \begin{tabular}{cccc}
    \toprule[1pt]
        Language & Dialect & Number of Contributors \\
        \midrule[0.5pt]
        English                & - & 39 \\
        \multirow{2}*{Chinese} & Mandarin & 121 \\
                               & Others & 46 \\
    \bottomrule[1pt]
    \end{tabular}
\end{table}

Each sample is recorded by a near-field mobile phone close to the speaker at around 0.2 meters and stored in a single-channel 16-bit stream with a sampling rate of 16kHz. We divided the dataset into four subsets according to different phases in the challenge, namely training, practice, feedback and private. The training set, recorded from 100 speakers, is used for participants to develop Auto-KWS solutions. The practice set contains data from five speakers, each with five enrollment audios and several test samples. Together with the downloadable docker image, the practice dataset provides an example of how the platform would call the participants’ code. To facilitate offline debugging, both training and practice sets are open for download. The feedback and private datasets with the same format as the practice set are for final evaluation and unavailable to participants. The platform will evaluate each participant’s final solution using those two datasets during the feedback and private phases, respectively, without any human intervention.

\begin{table}[H]
\caption{\textit{{\textbf{The summary of datasets} illustrates the data distribution in each subset. Each subset corresponds to a specific competition phase. Over half of the data are released to allow the participants to fine-tune their models better.}}}
\scalebox{0.81}{
    \begin{tabular}{cccc}
    \toprule[1pt]
        Dataset & Speaker Number & Phase & Enrollment Number \\ 
        \midrule[0.5pt]
        Training & 100 & Before feedback & 10 \\
        Practice & 5 & Before feedback & 5 \\
        Feedback & 20 & Feedback & 5 \\ 
        Private & 40 & Final & 5 \\
    \bottomrule[1pt]
    \end{tabular}
}
\end{table}

Test data in practice, feedback and private phases have the same format. For each speaker, only samples containing the wake-up word from the speaker are positive (wake-up word only at the end of audio). All other samples, such as other utterances from the speaker, the wake-up word recorded by other speakers, and any other audio that not contains the wake-up word recoreded by the speaker, are negative. In order to increase the number and diversity of test data, we have expanded the test data from the following directions:
\begin{itemize}
\item splicing two pieces of audio
\item adding the MUSAN noises \cite{snyder2015musan}, the signal-to-noise-ratio(SNR) was set between 5dB to 25dB
\item adding the RIRs-NOISES \cite{ko2017study}, the mixture weight was 0.5
\item perturbing the volume, the perturbation scale was set between 0.5 to 2
\end{itemize}
\subsection{Starting kit}
\label{subsec:starting-kit}

We provide the participants with a starting kit\footnote{https://github.com/janson9192/Auto-KWS2021}, which contains practice datasets, submission sample code, two baselines, and the ingestion and scoring code that has the similar call logic with the online challenge platform. Participants can create their own code submission by just modifying the dir ``code\_submission", which contains ``initialize.sh", ``enrollment.sh", and ``predict.sh", or adding other dependency code files including pretrained models. Then they can upload the zip-package of the submission folder.

It is very convenient to test and debug locally with the same handing programs and Docker image\footnote{https://hub.docker.com/r/janson91/Auto-KWS2021} of the Challenge platform, and evaluate by experimenting with practice datasets. Starting kit can be run in both CPU and GPU environment, but two baselines need GPU environment and the version of cuda cannot be lower than 10. Participants can check the python version and install python packages in the docker.

\section{Baseline and Experiments}
\label{sec:baseline-experiments}
\vspace{4mm}

\subsection{Baseline method 1}
\label{subsec:baseline-method-1}

\begin{figure}
    \centering
    \includegraphics[scale=.4]{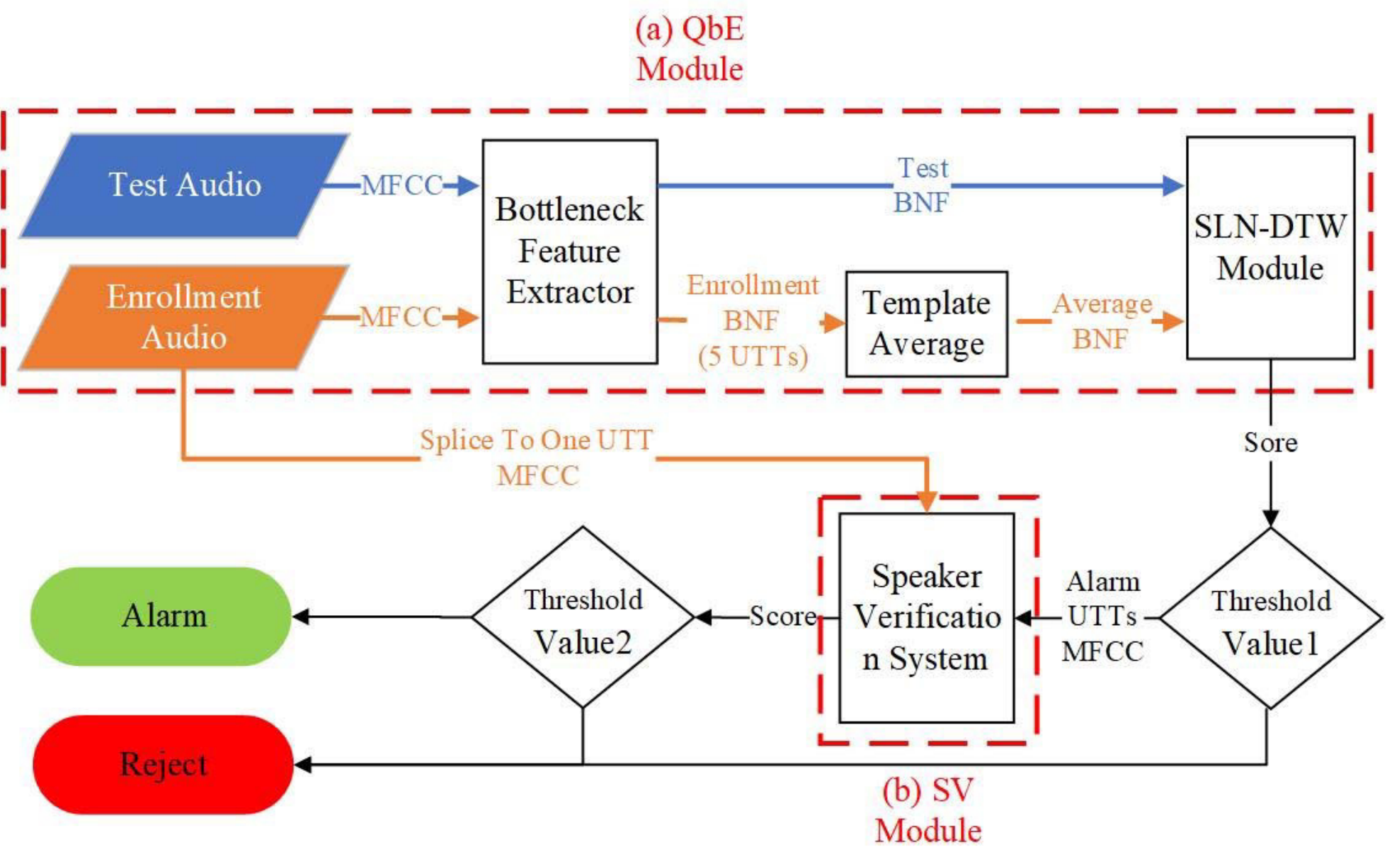}
    \caption{The system framework for baseline method 1. This method is composed of (a) query by example (QbE) module and (b) speaker verification (SV) module. The alarm utterances will be sent to SV module to re-verify.}
    \label{fig:baseline1_system}
\end{figure}

\begin{table}[ht]
    \caption{The network config of the BNF extractor}
    \centering
    \footnotesize
    \label{tab:BNF_extractor_config}
    \begin{tabular}{ccc}
    \toprule[1pt]
    \textbf{Layer}  & \textbf{Input Context Frame} & \textbf{Output Dim}      \\ \hline
    LDA    & -2, -1, 0, 1, 2     & 5 * 40          \\ \hline
    TDNN1  & 0                   & 850             \\ \hline
    TDNN2  & -1, 0, 2            & 850             \\ \hline
    TDNN3  & -3, 0, 3            & 850             \\ \hline
    TDNN4  & -7, 0, 2            & 850             \\ \hline
    TDNN5  & -3, 0, 3            & 850             \\ \hline
    BNF    & 0                   & 40              \\ \hline
    TDNN6  & 0                   & 850             \\ \hline
    output & 0                   & targets number  \\ \bottomrule[1pt]
    \end{tabular}

\end{table}
As shown in Fig.~\ref{fig:baseline1_system}, Baseline method 1 consists of a query by example (QbE) system and a speaker verification (SV) system. QbE is one of the text-independent wake-up systems, which we can use any speech as wake-up words without retraining models. This system consists of a bottleneck feature extractor, a template average module, and a segmental local normalized DTW (SLN-DTW) module. If the output score from SLN-DTW is bigger than the threshold value $\gamma_1$, we use the SV module to re-verify. Specifically, we splice the enrollment audios together to a long audio, and then by comparing the speaker vector similarity (by cosine distance) of the long audio and wake-up audios, we can decide whether to wake up finally. Of course, we need a threshold value $\gamma_2$ here.

We use Kaldi tools~\cite{povey2011kaldi} to train the bottleneck feature (BNF) extractor model in the QbE module. The BNF extractor is a special time-delay neural network (TDNN) automatic speech recognition (ASR) acoustic model. We use 40-dim MFCC features as input. The model structure is shown in Table \ref{tab:BNF_extractor_config}. It refers to \emph{/kaldi/egs/aishell2/s5/local/nnet3/run\_tdnn.sh}, and the difference is that we add a 40-dim affine-layer (we call it BNF layer) between TDNN5 and TDNN6 layer. The outputs of the bottleneck layer are the BNFs. We train this model by Magicdata dataset (\emph{http://www.openslr.org/68}) without data augmentation.

\begin{table*}[!t]
    \caption{\it \textbf{Baseline Results} on feedback and private datasets in Auto-KWS 2021. Datasets in feecback phase and final phase contains 20 speakers and 40 speakers respectivately. Average Score, Average Miss Rate and Average FA Rate are calculated based on the baselines' predictions of all utterances. Compute time represents the running time of the whole program, which contains initialization, enrollment and prediction processes. }
    \label{tab:baseline_results}
    \centering
    \setlength{\tabcolsep}{2mm}{
    \begin{tabular}{ccccccc}
    \toprule[1pt]
        Phase & Baseline & Average Score & Average Miss Rate & Average Fa Rate & Compute Time &  \\ \midrule[0.5pt]
        \multirow{2}*{Feedback} & Baseline 1 & 0.859 & 0.443 & 0.046 & 0:54:07 &  \\
         & Baseline 2 & 1.695 & 0.481 & 0.135 & 1:24:07 &  \\ \midrule[0.5pt]
        \multirow{2}*{Final} & Baseline 1 & 0.742 & 0.531 & 0.023 & 1:57:15 &  \\
         & Baseline 2 & 1.086 & 0.691 & 0.044 & 3:01:32 &  \\
    \bottomrule[1pt]
    \end{tabular}}
\end{table*}
\begin{table}[]
\centering
\footnotesize
\caption{The data augment config of the SV model}
\label{tab:SV_data_config}
\begin{tabular}{ccc}
\toprule[1pt]
\textbf{\begin{tabular}[c]{@{}c@{}}Augment\\  Type\end{tabular}} & \textbf{Detail}                                                    & \textbf{Times} \\ \hline
\multirow{2}{*}{reverberation}                                   & \begin{tabular}[c]{@{}c@{}}small room\\ (1m to 10m)\end{tabular}   & 0.5            \\ \cline{2-3} 
                                                                 & \begin{tabular}[c]{@{}c@{}}medium room\\ (10m to 30m)\end{tabular} & 0.5            \\ \hline
\multirow{2}{*}{noise}                                           & \begin{tabular}[c]{@{}c@{}}MUSAN noise\\ SNR: 15,10,5\end{tabular}  & 0.5            \\ \cline{2-3} 
                                                                 & \begin{tabular}[c]{@{}c@{}}MUSAN music\\ SNR: 20,15,10\end{tabular} & 0.5            \\ \hline
origin                                                           & ---                                                                & 1              \\ \hline
\multicolumn{2}{c}{total}                                                                                                             & 3              \\ \bottomrule[1pt]
\end{tabular}
\end{table}

The template average module and SLN-DTW module refer to \cite{hou2016investigating} (codes in \emph{http://www.openslr.org/68}). The two modules are developed in C++. We use the template average module to average the enrollment BNFs. And then use the SLN-DTW module to compare the similarity of the average BNF and the test BNF. The alarm threshold value $\gamma_1$ is set to 0.80.

At last, we train the X-Vector speaker verification module by Kaldi tools, too. The codes refer to \emph{/kaldi/egs/sre16/v2/run.sh}. We augment the Magicdata dataset three times to train this model, with details in Table \ref{tab:SV_data_config}. And the SV threshold value $\gamma_2$ is set to 0.83.

\subsection{Baseline method 2}
\label{subsec:baseline-method-2}

Baseline method 2 uses a very simple code pipeline, which contains a pretrained  model \cite{baevski2020wav2vec} for feature extraction, and a DTW \cite{giorgino2009computing} discriminator. 
The pretrained model \footnote{https://dl.fbaipublicfiles.com/fairseq/wav2vec/wav2vec\_small.pt} 
from the repo \footnote{https://github.com/eastonYi/wav2vec.git} is 
trained by Librispeech dataset \footnote{http://www.openslr.org/12} without finetuning, and we use the \textit{quantization module} to extract speech representation, with 256 dimensions.

After feature extraction, DTW is used for feature comparison. 
A long test audio will be compared with the enrollment audio segment by segment, and the minimum score will be taken as the score of the test audio. 
If the score is bigger than the threshold, the audio is judged to be awake. 
The awake threshold is calculated by DTW scores from the enrollment audio, and  constrained between 0.45 and 0.6.

\subsection{Experiments}
\label{subsec:experiments}
We submitted two baselines mentioned above to the challenge platform, according to the competition process. Each baseline was submitted twice and run on feedback and private datasets automatically.
The experiments are carried out on Google Cloud virtual machine instances under Ubuntu 18.04, with one single GPU (Nvidia Tesla P100) running CUDA 10 with drivers cuDNN 7.5, 100 GB  Disk and 26 GB Memory. The experiments are constrained to be completed within the same limited time as the requirement to participants.
The results on feedback and private datasets are presented in Table \ref{tab:baseline_results}, and details can be viewed on the platform.

As shown in Table \ref{tab:baseline_results}, the performance of Baseline method 1 is better than that of Baseline method 2 in terms of Miss Rate, False Alarm Rate and final Average Score. 
Moreover, the running efficiency of Baseline method 1 is higher than that of Baseline method 2, although the pipeline of Baseline method 2 is much simpler.
This challenge is a particularly difficult problem, so the performance of the two baselines is relatively poor, leaving a lot of room for participants to improve.

\section{Conclusions}
\label{sec:conclusion}
\vspace{4mm}

Auto-KWS 2021 focuses on Automated Machine Learning for customized Keyword Spotting tasks. In this paper, we introduce the setup of the Auto-KWS 2021 and describe the protocol, metrics, datasets,  starting kit and two baseline systems of the challenge. 
Personalized and customizable KWS is a very interesting and practical scene. According to this baseline experiment, however, it is still a field that has not been conquered. It is expected that researchers from both academia and industry can advance problem solving through this challenge.

\section{Acknowledgements}
\label{sec:acknoledgements}
\vspace{4mm}
This project was supported in part by 4Paradigm Inc., ASLP@NPU and ChaLearn.
The authors were grateful to Isabelle Guyon and Zhen Xu for computational resources.
The platform, automl.ai\footnote{https://www.automl.ai}, is built based on Codalab\footnote{https://competitions.codalab.org}, an web-based platform for machine learning competitions.
Thanks to Xiawei Guo, Shouxiang Liu and Zhen Xu for their support on challenge platform construction.

\clearpage

\bibliographystyle{IEEEtran}

\bibliography{refs}

\end{document}